\documentclass[twocolumn, pre, showpacs,preprintnumbers,amsmath,amssymb, showkeys]{revtex4}
\usepackage{graphicx}
\usepackage{dcolumn}
\usepackage{bm}
\usepackage{hyperref}

\begin{document}

\title{A Random Loop Model for Long Polymers}

\author{Manfred Bohn}
\email{bohn@tphys.uni-heidelberg.de}
\author{Dieter W. Heermann}
\affiliation{Institute of Theoretical Physics, University of Heidelberg, Philosophenweg 19, D-69120 Heidelberg, Germany}
\author{Roel van Driel}
\affiliation{Swammerdam Institute for Life Sciences, University of Amsterdam, BioCentrum Amsterdam, Kruislaan 318, 1098 SM Amsterdam, The Netherlands}

\date{\today}

\begin{abstract}
Remarkably little is known about the higher-order folding motifs of the chromatin fibre inside the cell nucleus. Folding depends among others on local gene density and transcriptional activity and plays an important role in gene regulation. Strikingly, at fibre lengths above 5 to 10 Mb the measured mean square distance $\left<R^2\right>$ between any two points on the chromatin fibre is independent of polymer length. We propose a polymer model that can explain this levelling-off by means of random looping. We derive an analytical expression for the mean square displacement between two arbitrary beads. Here the average is taken over the thermal ensemble with a fixed but random loop configuration, while quenched averaging over the ensemble of different loop configurations -- which turns out to be equivalent to averaging over an ensemble of random matrices --  is performed numerically. A detailed investigation of this model shows that loops on all scales are necessary to fit experimental data.
\end{abstract}
\maketitle

\section{Introduction}
The folding behaviour of DNA inside the cell nucleus has been subject to extensive studies.  The diploid human genome consists of about 2 m of double stranded DNA, which is wrapped around histone protein octamers, forming nucleosomes every 180 to 250 bp. This chromatin fibre has a length of about 50 cm and is packaged inside a cell nucleus of about 10 $\mu$m diameter. Despite its tight folding, the DNA is readily accessible to large numbers of proteins required for transcription, replication and DNA repair. The first stages of compaction are well-known~(see ref.~\cite{Schiessel2001} for a review): The DNA double strand is wrapped around histone octamers in a 1-and-3/4 left-handed superhelical turn forming the so-called nucleosomes, which are connected by stretches of linker DNA. This nucleosome-linker complex has a diameter of about 10 nm and is called 10 nm chromatin fibre. It has been shown \textit{in vitro} that this fibre in turn condenses to a fibre of 30 nm diameter, whose structure is still under discussion~\cite{Widom1989, Holde1995, Holde1996}. 

Remarkably little is known about the higher-order folding of the chromatin fibre inside the human interphase nucleus~\cite{Horowitz-Scherer2006}. It is not possible to follow the chromatin fibre in the interphase nucleus by imaging techniques. Therefore indirect approaches are being used to study chromatin folding. One method that has been applied by several groups is fluorescence in situ hybridization (\textit{FISH}). By labelling chromatin in the intact cell with pairs of \textit{FISH}-markers one can measure the mean squared physical distance $\left<R^2\right>$ between these markers as a function of the genomic distance $d$. A recent study has shed light on the folding at larger length scales~\cite{Mateos2007}. It was shown that there is a levelling-off in the physical vs. genomic distance plot at genomic distances longer than a few mega base pairs (Mb), so that approximately $\left<R^2\right> \sim O(1)$ for large  $d$. At shorter length scales, i.e. up to a few Mb, the folding behaviour is best described within the globular state model, where the mean square physical distance scales like
\begin{equation}
 \left<R^2\right> \sim d^{2/3}\qquad .
\end{equation}
The unexpected behaviour at larger genomic distances called for further explanation. 

In the past, there have been several attempts to explain the folding motifs of chromatin in interphase cells using polymer models~\cite{Hahnfeldt1993, Yokota1995, Sachs1995, Munkel1998}. One approach has been to model the chromatin fibre as a random walk in confined geometry~\cite{Hahnfeldt1993}. Other data have been interpreted in terms of a Random-Walk/Giant-Loop (RWGL) model suggesting that there is a random walk backbone with regularly placed loops of about 3 Mb~\cite{Yokota1995, Sachs1995} to explain different observed folding regimes on large and small scales. Another model, the Multi-Loop-Subcompartment (MLS) model~\cite{Munkel1998, Munkel1999}, proposes the existence of a rosette-like structure with 120 kb loops. None of these models presented so far is able to explain the levelling-off observed in the recent study.

It has been shown that the formation of loops of different sizes plays an important role in gene regulation and gene expression. One system that has been extensively studied is the $\beta$-globin locus~\cite{Palstra2003, Laat2003}. Here chromatin expression is controlled by the formation of loops bringing different regulatory elements of the locus in physical contact. In this system the loop sizes are in the order of 10 kb. The formation of these loops is dynamic: different genes in the locus interact with the control locus during development in a mutually exclusive way, correlated with their expression. Loops that link promoter and enhancer complexes have been found of up to 3 Mb~\cite{Petrascheck2005, Fraser2006}. Loops of up to several tens Mb have been associated with the formation of transcription factories, which bring together transcriptionally active genes ~\cite{Fraser2006, Fraser2007}. In all cases, chromatin loop formation is a dynamic process~\cite{Cook2002}.

\begin{figure}
 \includegraphics[angle=270, width=\hsize]{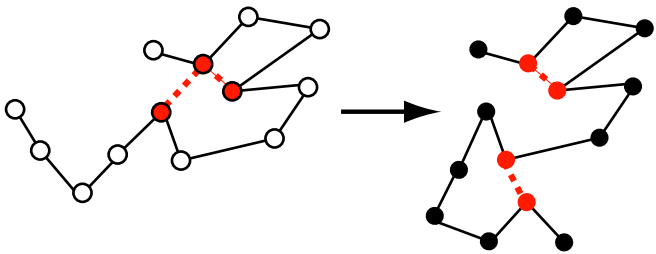}
\caption{\label{fig:configurations} The Random Loop Model averages over (a) the thermal disorder and (b) over the possible configurations of loops. Here one can see two possible configurations of loops. }
\end{figure}

Based on these observations we propose a general polymer model which is able to explain the levelling-off behaviour observed in experiment where the mean squared physical distance scales like $\left<R^2\right> \sim O(1)$ at genomic distances above 1-2 Mb. The model takes into account the looping of the polymer, i.e. the chromatin fibre. The backbone of our polymer is formed by a random walk chain. For reasons of mathematical tractability we do not introduce excluded volume interactions. While Sachs \textit{et al.} assumed loops of uniform size and fixed positions of the loop attachment points~\cite{Sachs1995}, we allow loops to have random polymer length and attachment points.  We are interested in the average conformational properties of the system, mainly the mean square distance between two beads of the chain. We derive an analytical result for the average over the thermal disorder of one specific configuration of loops. The randomness of the loops gives rise to average over all different loop configurations. Numerical methods are used to get the quenched average over the different configurations of loops (see fig.~\ref{fig:configurations} for two possible configurations).  This average corresponds to the cell-to-cell variation of the measurements. We compare our model to the experimental data and investigate which loops are necessary to get a levelling-off where $\left<R^2\right> \sim O(1)$. 

\section{Theory}
\subsection{Polymer models}
In the past, experimental data from \textit{FISH} measurements has been compared quite successfully to polymer models. Mainly three basic polymer models are used: (a) the random walk (RW) model where it is assumed that the orientations between chain segments are completely uncorrelated and no volume interactions are taken into account, (b) the self-avoiding walk model (SAW) takes excluded volume into account, while (c) the globular state (GS) model furthermore includes temperature-dependent attractive interactions (cf. ref.~\cite{Grosberg1994, Gennes1979}). One characteristic feature of a polymer model is the mean squared end-to-end distance $\left<R_N^2\right>$. For the basic models above, the end-to-end distance scales like
\begin{equation}\label{eq:scalinglaw}
 \left<R_N^2\right> = b^2 N^{2\nu}
\end{equation}
in the limit of large $N$, where $l$ is the linker length, $N$ the chain length and $\nu$ a constant depending on the model used: $\nu=0.5$ for the RW, $\nu\approx 0.588 $ for the SAW and $\nu=1/3$ for the GS. Assuming that eq.~\eqref{eq:scalinglaw} also holds for intra-chain distances in the limit of large $N$, one can compare the experimental data to these models~\cite{Mateos2007}. Relation~\eqref{eq:scalinglaw} is no longer valid as soon as we introduce long-range interactions along the chain, such as chromatin looping.

Polymer modelling makes use of a coarse-graining approach, i.e. we divide the chromatin fibre into $N$ equal subunits of length $b$. The subunits itself are assumed to be uncorrelated, i.e. they can freely rotate around each other. Such a description of a biological polymer is correct when we make $N$ sufficient small so that $b$ is larger than the persistence length of chromatin, i.e., the bending energy vanishes on this length scale. 

\subsection{General expression for the average over the thermal ensemble}
\label{sec:theory:genexp}
To set up our Random Loop Model we first derive a general expression for the mean square distance between two arbitrary beads of the chain where harmonic interactions between all beads of the chain are allowed. The spatial positions of the chain's beads are denoted by $\mathbf{x}_0, \ldots, \mathbf{x}_N$ and $\mathbf{x}_i = (x_i, y_i, z_i)$. For reasons of mathematical tractability we consider a phantom chain (random walk) where no excluded volume is taken into account. The Gaussian chain that forms the backbone of our model  is characterized by the following potential, 
\begin{equation*}
U_{\text{Gaussian}} =  \frac{\kappa}{2} \sum_{j=1}^{N} \parallel\mathbf{x}_j - \mathbf{x}_{j-1}\parallel^2\;.
\end{equation*}
In addition to the random walk backbone, we allow each bead to interact with any other via harmonic potential, resulting in an interaction term in the potential
\begin{equation*}
 U = U_{\text{Gaussian}} + \frac{1}{2}\sum_{\substack{i < j\\|i-j|>1}}^N \kappa_{ij} \parallel\mathbf{x}_i - \mathbf{x}_j\parallel^2
\end{equation*}
where $\kappa_{ij} = \kappa_{ji}$ are the spring constants for the loop attachment points. The spring constants $\kappa_{ij}$ are given here and in the following in reduced units (comprising the term $1/k_BT$).  Right now we keep them arbitrary but they will be randomly chosen later within the model. This potential has already been proposed by Sachs \textit{et al.}~\cite{Sachs1995} but not been elaborated further. It can be rewritten in the form
\begin{equation} \label{eq:general:potential}
 U = \frac{1}{2} \sum_{i < j}^{N} \kappa_{ij} \parallel \mathbf{x}_i - \mathbf{x}_j \parallel ^2 = \frac{1}{4} \sum_{\substack{i,j=0\\j\neq i}}^{N} \kappa_{ij} \parallel \mathbf{x}_i - \mathbf{x}_j \parallel ^2 \;.
\end{equation}
where $\kappa_{ij}=\kappa$ for $|i-j|=1$.

The probability density for a bead conformation $(\mathbf{x}_0, \ldots, \mathbf{x}_N)$ in the canonical ensemble is given by the Boltzmann factor
\begin{equation} \label{eq:general:G}
P(\mathbf{x}_0, \ldots \mathbf{x}_N) = C \: \exp\left( -\frac{U}{k_B T}\right)\quad, 
\end{equation}
where $C$ is a normalization constant and $U=U (\mathbf{x}_0, \ldots \mathbf{x}_N)$ is the total potential energy of the chain. 

We now eliminate the degrees of freedom stemming from the translational invariance of the problem by setting $\mathbf{x}_0 = \mathbf{0}$ (the absolute position of the chain in space is irrelevant for distances between beads). 

Due to the Gaussian character of the probability density $G$, the spatial dimensions factorize,
\begin{multline*}
P(\mathbf{x}_1, \ldots, \mathbf{x}_N) = \\ P_1(x_1, \ldots, x_N) \cdot P_1(y_1, \ldots, y_N) \cdot P_1(z_1, \ldots, z_N)
\end{multline*}
and we can concentrate on the one-dimensional density function $P_1$. By an easy calculation omitted here, we can rewrite  the one-dimensional probability  density $P_1$ as follows
\begin{equation}\label{eq:G1}
P_1(x_1, \ldots, x_N) = C_1 \exp(-\frac{1}{2} \mathbf{X}^T K \mathbf{X})
\end{equation}
where $\mathbf{X} = (x_1, \ldots, x_N)^T$ and 

\begin{equation}\label{eq:matrixK}
K = 
\begin{pmatrix}
\sum_{\substack{j=0\\j\neq 1}}^N \kappa_{1j} &  -\kappa_{12} & \dots & -\kappa_{1N} \\
-\kappa_{21} & \sum_{\substack{j=0\\j\neq 2}}^N \kappa_{2j} &  \dots & -\kappa_{2N} \\
\vdots & \vdots & \ddots & \vdots\\
-\kappa_{N1} & -\kappa_{N2} & \dots & \sum_{\substack{j=0\\j\neq N}}^N \kappa_{Nj} 
\end{pmatrix}
\end{equation}
Up to now we have not made any assumptions concerning the spring constants (i.e. basically the matrix entries) $\kappa_{ij}$. In the following we only assume that $K$ is a \textit{symmetric} and \textit{regular} matrix. 
So $P_1$ in eq.~\eqref{eq:G1} turns out to be a multivariate normal distribution with mean $\mathbf{\mu}=0$ and covariance matrix $\Sigma = K^{-1}$. The marginal distribution for two arbitrary beads $I$ and $J$, 
\begin{equation}
 P(x_I, x_J) = \int\ldots\int \prod_{\substack{i=1\\i\neq I, J}}^N P(x_1, \ldots, x_N)
\end{equation}
can be evaluated by standard methods for normal distributions. Going back to three dimensions we obtain after some basic integral evaluations the joint probability density for the distance between two beads $I$ and $J$,
\begin{multline*}
P(\parallel\mathbf{x}_I - \mathbf{x}_J\parallel) \equiv P(r_{IJ})=\\ \tilde{C}\, r_{IJ}^2\,\exp\left[-\frac{1}{2} \frac{1}{\sigma_{JJ} + \sigma_{II} - 2\sigma_{IJ}} r_{IJ}^2\right]\;.
\end{multline*}
Here
\begin{equation*}
\Sigma = K^{-1} = \left(\sigma_{ij}\right)_{i, j}
\end{equation*}
and $\tilde{C}$ is the normalization constant. Using 
\begin{equation*}
\Gamma = \frac{1}{2} \frac{1}{\sigma_{JJ} + \sigma_{II} - 2\sigma_{IJ}}
\end{equation*}
and calculating the correct normalization we obtain
\begin{equation} \label{eq:Pr}
P(r_{IJ}) = \frac{4}{\sqrt\pi}\; \Gamma^{\frac{3}{2}}\; r_{IJ}^2\; \exp\left[ -\Gamma r_{IJ}^2\right]
\end{equation}
and finally
\begin{eqnarray}
\left< r_{IJ}^2 \right>_{\text{thermal}} &=& \int r_{IJ}^2 P(r_{IJ}) dr_{IJ} \nonumber \\
					 &=& \frac{3}{2\Gamma} = 3 (\sigma_{JJ}+\sigma_{II} - 2\sigma_{IJ})\label{eq:r2} \;.
\end{eqnarray}

The bracket delimiters here denote the average over the thermal ensemble of $N+1$ beads interacting via a given, but fixed harmonic potential.
 
\subsection{The Random Loop Model}
The last section was dedicated to the derivation of a general formula for the mean square displacement between two arbitrary beads of the chain where each bead may interact with any other via harmonic potential. This quantity turned out only to depend on the matrix $K$, or more accurately speaking, on its inverse. The matrix $K$ contains all information about the interactions.  Now we want to specify this matrix.  Our model assumes the chromatin fibre to have a random walk backbone, meaning that $\kappa_{ij} = \kappa$ with $|i-j|=1$. Furthermore the chromatin forms loops whose size and positions are randomly distributed along the chain. On a more general footing we can restrict the possible loop sizes $\ell$ to a certain range $[l_1, l_2]$. Within this range all loops are chosen randomly by setting:
\begin{align*}
 \kappa_{ij} &= \begin{cases}
			\kappa\quad&\text{with probability }\mathcal{P}\\0\quad&\text{with probability }1-\mathcal{P}
		\end{cases}
, &\quad& \text{if } l_1 \leq |i-j| \leq l_2\\
\kappa_{ij} &= 0&\quad& \text{otherwise}
\end{align*}
Note that we can set $\kappa=1$ as it only scales the mean square displacement in eq.~\eqref{eq:r2}. Thus our model has two adjustable parameters, namely the chain length $N$ and the probability $\mathcal{P}$.

The resulting matrices ${K}$ represent an ensemble of diagonally dominated band random matrices and each matrix of this ensemble represents a loop configuration. This ensemble of random matrices has been investigated recently~\cite{Heermann2007}. We are interested in the ensemble average of the mean square displacement, i.e. in the quantity
\begin{eqnarray*}
\left<r_{IJ}^2\right> &=& \left< \left< r_{IJ}^2\right>_{\text{thermal}}\right>_{\text{loops}} \\
&=& 3 \left(\left<\sigma_{JJ}\right>_{\text{loops}}+\left<\sigma_{II}\right>_{\text{loops}} - 2\left<\sigma_{IJ}\right>_{\text{loops}}\right)\qquad .
\end{eqnarray*}
This average is a quenched average and is equivalent to averaging over the ensemble of random matrices given by the above constraints. In sec.~\ref{sec:ann} we also consider the case of the annealed ensemble and give an explanation why we use the quenched one here. 

As our model already assumes that the chromatin fibre is translational invariant (as we do not take into account genomic content), we are only interested in the mean square displacement $\left<R_n^2\right>$ for two beads separated by $n=|i-j|$. 

The average over the ensemble of random matrices cannot be performed analytically, so we have to use a representative subset of the ensemble and numerically calculate the inverse matrix and thereby the mean square displacement.
 
\section{Results}
\subsection{Comparison to experimental data}
As noted earlier, our polymer model makes use of coarse-graining, since it is impossible to model such a long fibre in detail. Restrictions are given by computing time, which basically depends on the size of the matrix $K$. For our calculations we chose a matrix size of $N=1000$ as a good compromise between computing time and not too coarse graining. 
Using a coarse-graining approach implies that we neglect details on a scale below the effective segment length being  150 kb in the following figures. Therefore we cannot resolve those loops that have been investigated in some gene-expression systems like the $\beta$-globin locus.  As we are interested in large scale chromatin organization, it is justified to neglect these loops as they have no effect for the levelling-off at large genomic distances, but only lead to a rescaling of the effective Kuhn length.

\begin{figure}
 \includegraphics[angle=270, width=0.95\hsize]{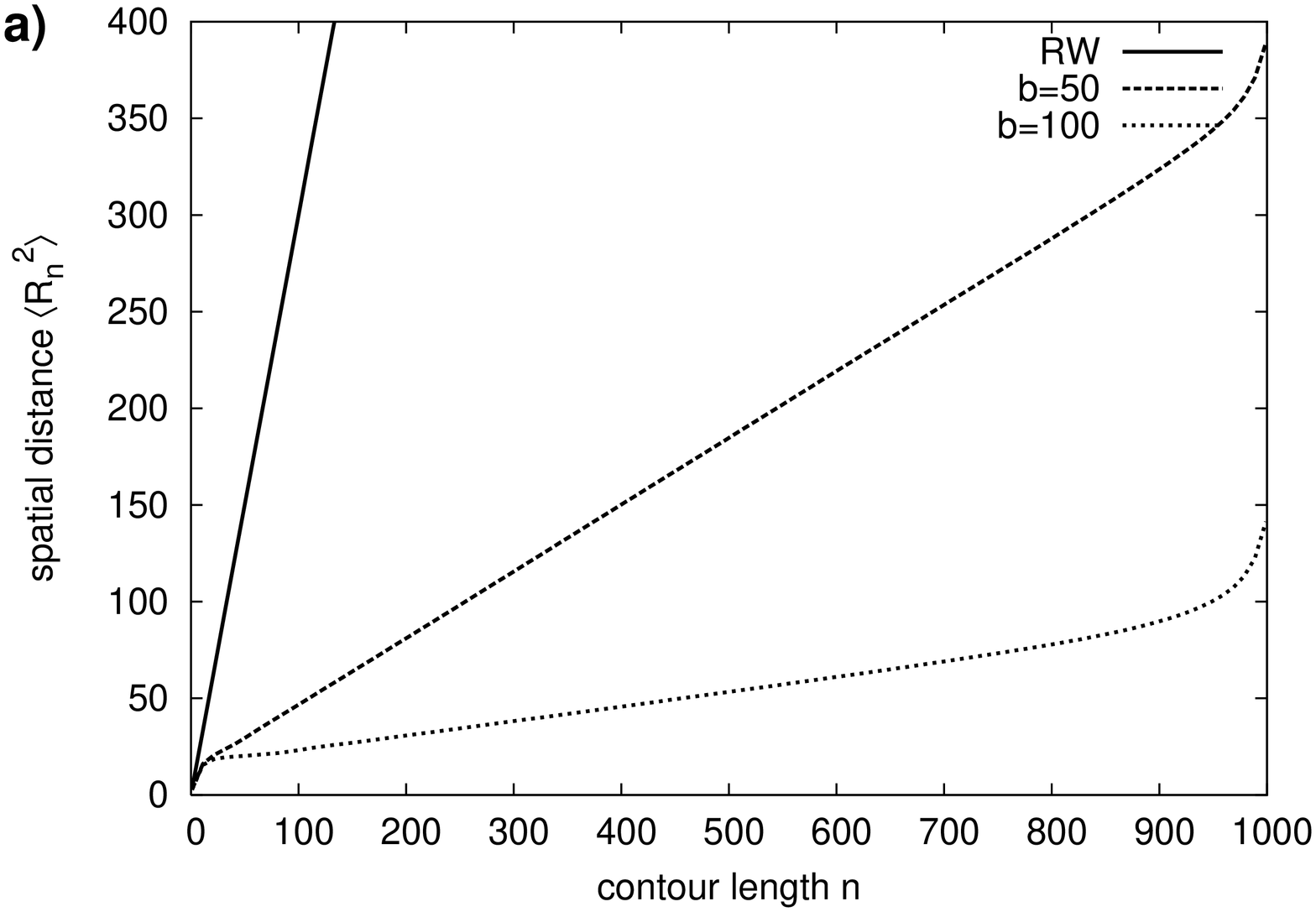}
\includegraphics[angle=270, width=0.95\hsize]{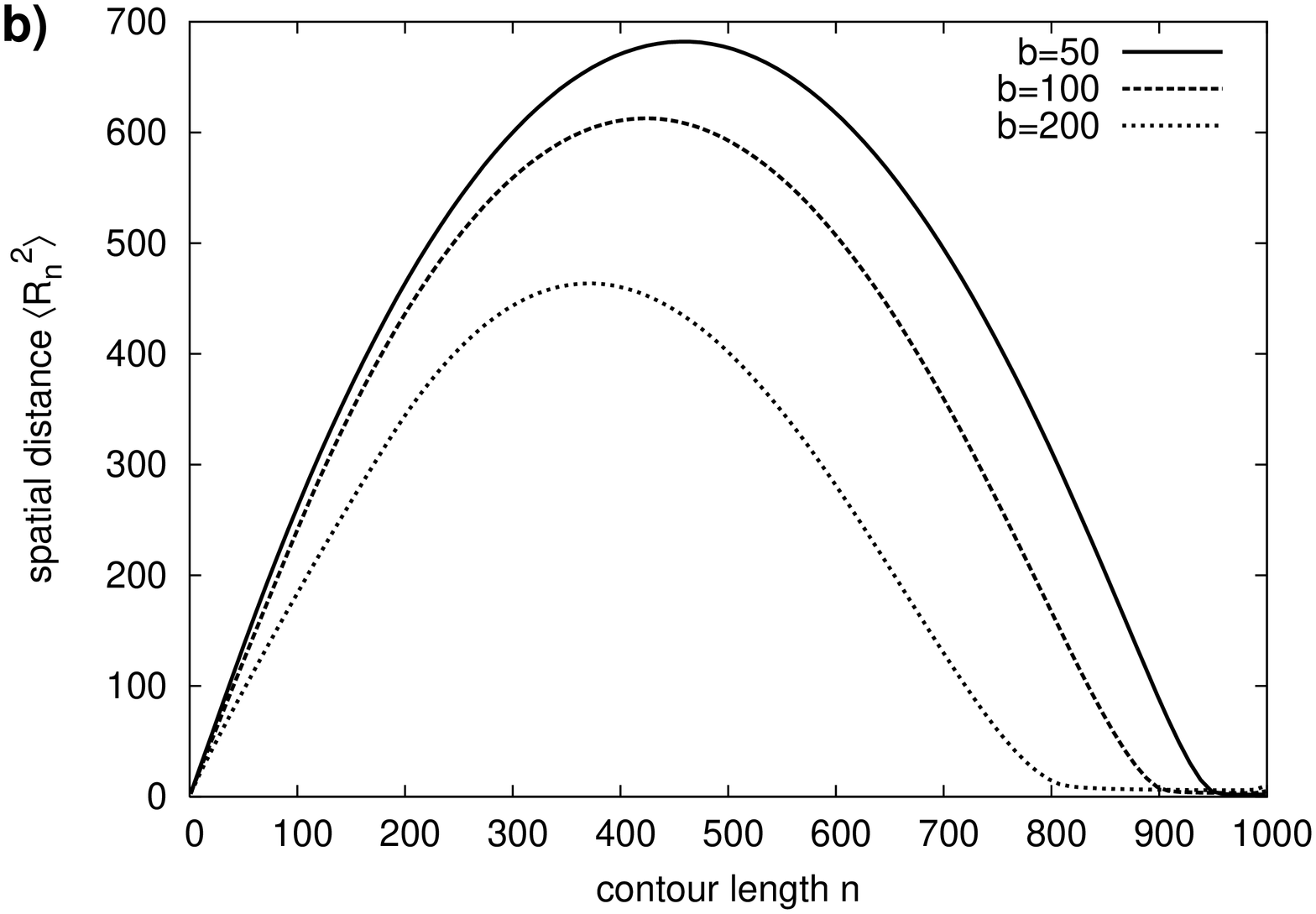}
\includegraphics[angle=270, width=0.95\hsize]{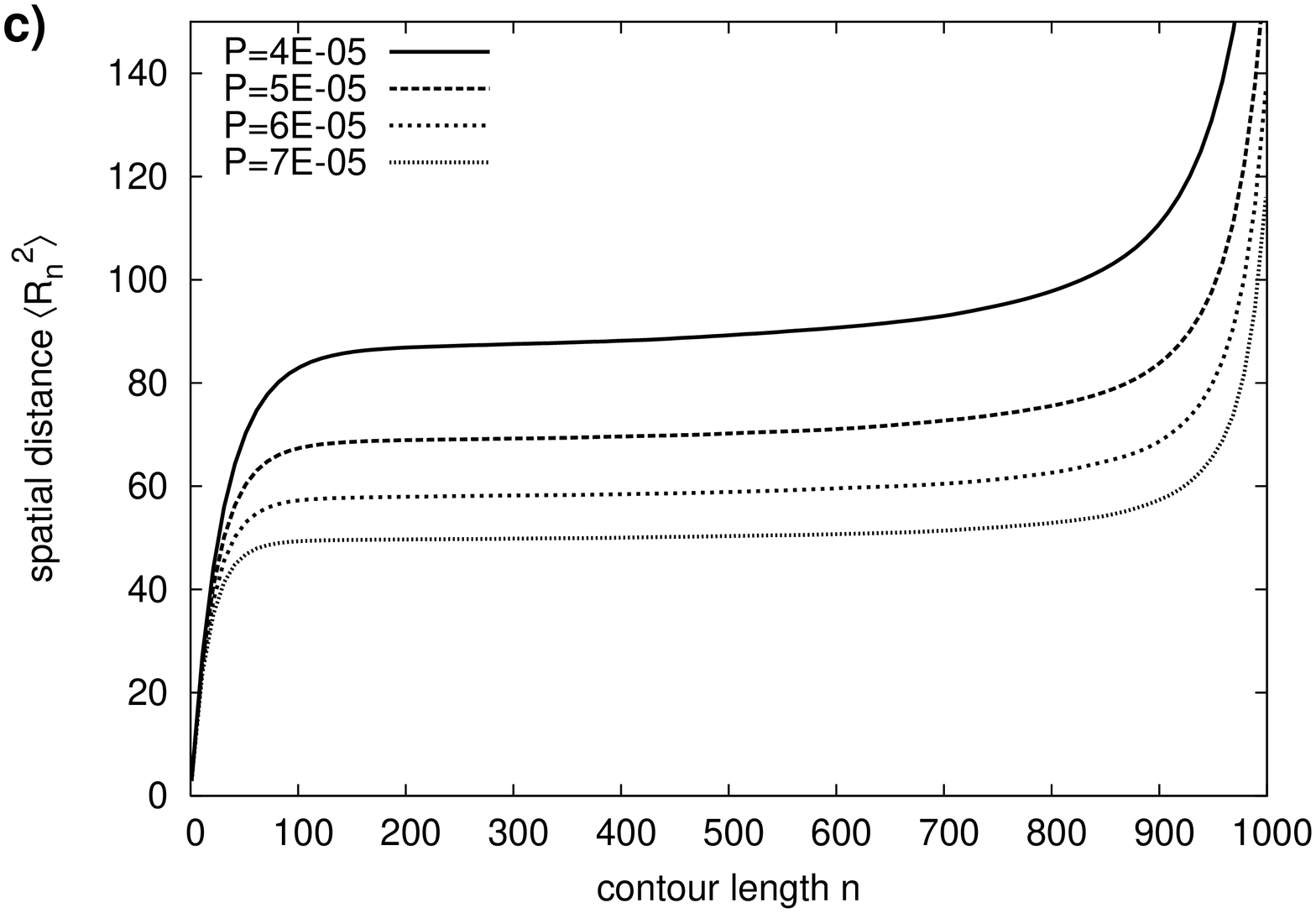}
\caption{\label{fig:loops} Mean square displacement between two chain segments  in relation to their contour length for the Random Loop Model for different allowed loop sizes. The chain length is always $N=1000$ In figure \textbf{a)} only loops smaller than a certain size $s$ are allowed. The basic scaling behaviour is still  $\left<R_n^2\right>\sim n$ with a changed effective contour length compared to the free random walk. For this plot $\mathcal{P}$ was chosen that the average number of loops per configuration is $100$. Figure \textbf{b)} is for large loops where only loops of sizes $\ell$ in a range $[N-s, N]$ are allowed. While large loops seem to be responsible for the collapse of the chain, they alone cannot explain the experimental data. As in a) $\mathcal{P}$ was chosen that the mean number of loops per configuration is $100$. Figure \textbf{c)} shows the results for the situation where loops of all sizes are allowed. The levelling-off to $\left<R_n^2\right> \sim O(1)$ can already be achieved by a small number of loops}
\end{figure}

The first point of interest is which loops are necessary for the observed experimental behaviour. Do small loops (in the order of 100 kb to 1 Mb) already lead to the levelling-off, or are loops on all scales up to 80 Mb needed? 
Restricting the loop sizes to a range $\ell \in [1, s]$ only leads to a rescaling of the effective segment length, we still have $\left<R_n^2\right>\sim n$ (fig.~\ref{fig:loops}a). 
In fig.~\ref{fig:loops}b) we analyzed the ensemble where only large loops are allowed. Obviously, large loops are responsible for forcing the collapse of the chain, but the overall behaviour of the mean square physical distance does not fit the experimental data. Therefore loops on all scales are needed to obtain the levelling-off observed in experiment.

The characteristic features of the mean square displacement allowing loops on all scales can be seen in fig.~\ref{fig:loops}c). At short contour lengths the mean square displacement grows similar to a random walk, but soon a levelling-off can be observed due to the attractive long-range interactions which is fairly $\sim O(1)$. While the contour length approaches $N$, the mean square displacement again rises to a random walk like behaviour. This is a chain end effect which is not of interest to us, as experiments only measure intra-chain distances. It is due to the construction of the loops, as the probability for having a loop with a larger size becomes increasingly small.

Thus, adding long-range interactions forcing the polymer to form loops yields completely different traits than a simple random walk or self-avoiding walk model. 
Note that the probabilities $\mathcal{P}$ are chosen very small, meaning that a few loops suffice to obtain this levelling-off. The number of independent randomly choosen entries $\kappa_{ij}$ is $\mathcal{C} = (N-1)(N-2)/2$ for a $N\times N$-matrix and therefore the average number of loops per configuration is given by $\mathcal{C}\cdot\mathcal{P}$. With $\mathcal{P}=4\times 10^{-5}$ and $N=1000$ one has an average of about $20$ loops. 

\begin{figure}
 \includegraphics[angle=270, width=\hsize]{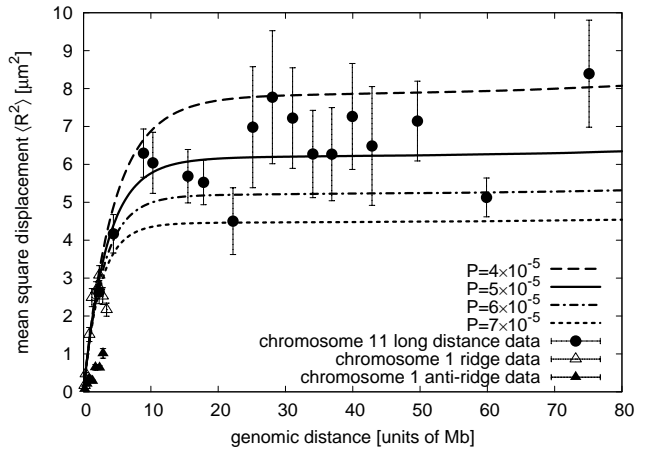}
\caption{\label{fig:exp}  Experimental data compared to the Random Loop Model. The data is taken from~\cite{Mateos2007} and includes short and long distance sets. The results of the Random Loop Model are shown for $N=1000$ and different values of $\mathcal{P}$. }
\end{figure}

In fig.~\ref{fig:exp} the model is compared to the experimental data for different values of $\mathcal{P}$. Here one has to introduce two new scaling parameters, the segment length in physical units (e.g. nm) and the segment length in base pairs. The data is shown for a segment length of $300$ nm and $150$ kb. The latter is the size of the flourescent markers used in experiments, therefore it does not make any sense to model on a more detailed scale. As mentioned above, using this coarse-graining approach all details on length scales smaller than 150 kb are neglected. The model can quite well explain the levelling-off at genomic distances above a few mega-base pairs as well as the rise at small genomic distances. As we have shown that on small genomic distances we have a globular-state-like behaviour~\cite{Mateos2007}, this random-walk-based model does not yield perfect results here. 

\begin{figure}
 \includegraphics[angle=270, width=\hsize]{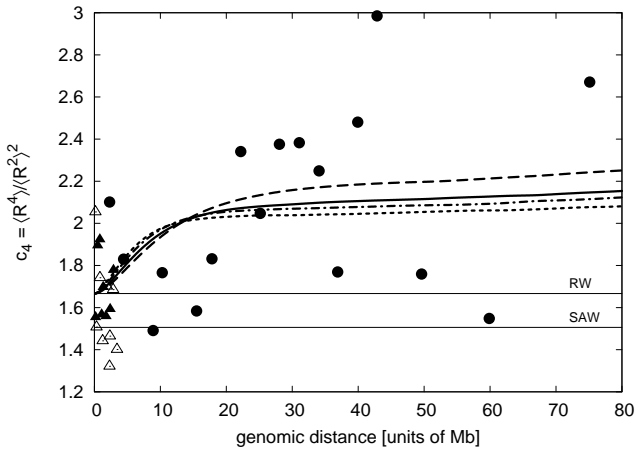}
\caption{\label{fig:cumulant} The cumulant expression $c_4 = {\left<R^4\right>}/{\left<R^2\right>^2}$ of the experimental data compared to SAW, RW (straight lines) and Random Loop Model (dashed lines). For the legend, see fig.~\ref{fig:exp}. The data shows significant differences to a RW and a SAW (horizontal lines), and is fluctuating  around the cumulant of the Random Loop Model.}
\end{figure}

In a recent publication~\cite{Mateos2007} we already mentioned that plotting $\left<R_n^2\right>$ versus $n$ is not a very sensitive method to check for the correctness of a model. Looking at the cumulant relation between higher-order moments,
\begin{equation}\label{eq:cumulant}
 c_4 = \frac{\left<R^4\right>}{\left<R^2\right>^2}
\end{equation}
 gives much stronger evidence, as this expression is related to the distribution of the distances and not only its average value. Furthermore it has the advantage that the physical length scale divides out. Eq.~\eqref{eq:cumulant} can be easily evaluated for a Gaussian Chain, where $c_4^{\text{RW}} = 5/3$. To obtain the value of the cumulant relation for a self-avoiding walk one has to use the expression for the distance probability density obtained by scaling arguments~\cite{Baiesi2003, Fisher1966}, 
\begin{equation}\label{eq:SAW:P}
 P_{\text{SAW}}(R_N) = A R_N^{\mu+2}\:\exp\left(-D\: R_N^{\frac{1}{1-\nu}}\right), \quad \mu=0.28
\end{equation}
Numerical integration gives $c_4^{\text{SAW}} \approx 1.506$, so both RW and SAW yield a constant expression. Fig.~\ref{fig:cumulant} shows that this constant has a value significantly below the fluctuations of the data. Here our model is in better agreement with experiments. We should point out here the importance of averaging over the disorder of loops. The cumulant expression $c_4$ only averaged over the thermal ensemble given by equations~\eqref{eq:Pr} and \eqref{eq:r2} is the same as for a pure random walk, namely $5/3$. It is the average over the loop configurations that changes this behaviour, bringing it in better agreement with the data. 

\subsection{Annealed vs. quenched average}
\label{sec:ann}
In systems with disorder one has to perform averages both over a set of statistical variables and over a set of random variables representing the disorder~\cite{Binder1986}. 
In the case of the Random Loop Model the partition sum $\mathcal{Z}$ depends on the statistical variables $\mathbf{x}_1, \ldots, \mathbf{x}_N$ and a set of random variables representing the disorder $\{\kappa_{ij}\}$ with $|i-j|>1$.
Above we have performed the average first over the statistical variables, while the average over the disorder has been taken over the quantity $\left<r_{ij}\right>^2$. This corresponds to a process of quenched averaging. Using this method we assume that the time the cell needs to go into a new disorder configuration is much longer than the time needed for the cell to come into thermal equilibrium, $\tau_{\text{eq}} \ll \tau_{\text{dis}}$. At least to our knowledge, such time scales are not known inside the cell nucleus. Time evolution measurements cannot be performed on a cell using \textit{FISH} markers as the cell has to be fixated before applying imaging techniques. Different configurations can only be observed by looking at different cells.  It seems to us more reasonable to use a quenched type of average for comparison to biological data, as the loops are functional complexes which have to persist a while in order to properly fulfill their tasks. 

Nevertheless, it is interesting to consider the other case, the annealed ensemble. This type of average should be applied if $\tau_{\text{dis}} \ll \tau_{\text{eq}}$. The annealed average of the partition sum can be written as
\begin{equation}
 \left<\mathcal{Z}\right>_{\text{ann}} = \sum_{\{\kappa_{ij}\}} \mathcal{Z}\left( \{\mathbf{x}_k\}, \{\kappa_{ij}\}\right)\, p\left(\{\kappa_{ij}\}\right)\quad, 
\end{equation}
where the sum is over all possible configurations of disorder, and $p$ is the probability of one such configuration. We restrict our calculation to the case that loops of all sizes are allowed and that the spring constant is $\kappa$ for adjacent beads as well as for loops. Assuming that the $\kappa_{ij}$ are i.i.d. Bernoulli as before, the average over the disorder can be carried out exactly, 
\begin{multline*}
 \left<\mathcal{Z}\right>_{\text{ann}} = \int d\mathbf{x}_1\ldots d\mathbf{x}_N \exp(-U_\text{Gaussian}) \\ \times  \prod_{i<j-1} \left[ \mathcal{P}\left(e^{-\frac{1}{2}\kappa \parallel \mathbf{x_i} - \mathbf{x}_{j}\parallel^2}-1\right) + 1\right]\quad .
\end{multline*}
Introducing the effective potential
\begin{equation*}
U_{\text{eff}} =  \frac{1}{2}\kappa\sum_{i=0}^{N-1}r_{i,i+1}^2  - \sum_{|i-j|>1} \log\left[ 1 + \mathcal{P} \left( e^{-\frac{1}{2}\kappa r_{ij}^2} - 1\right)\right]\;, 
\end{equation*}
where $r_{ij} = \parallel \mathbf{x}_i - \mathbf{x}_j\parallel ^2$,  we can rewrite the partition sum as
\begin{equation}\label{eq:Zann:Ueff}
 \left<\mathcal{Z}\right>_{\text{ann}} = \int  \int d\mathbf{x}_1\ldots d\mathbf{x}_N \exp(-U_\text{eff})	
\end{equation}
The effective potential has two parts: Adjacent beads with $|i-j|=1$ keep their attractive harmonic potential, while \textit{all} non-adjacent beads interact via a pairwise attractive potential $V(r)$. This potential is characterized by a minimum at $r=0$, while for large $r$ it reaches a plateau at $V(r \rightarrow \infty) = -\log(1-\mathcal{P})$. In a low temperature approximation, a series expansion around $r=0$ up to second order gives 
\begin{equation}\label{eq:V:ann:lt}
V(r) = \frac{1}{2}\mathcal{P}\kappa r^2 
\end{equation}
 -- a harmonic potential with effective spring constant $\mathcal{P}\kappa$.  

The partition sum in~\eqref{eq:Zann:Ueff} cannot be evaluated analytically and therefore we do not obtain an expression for the mean square displacement in the annealed case. One could obtain results using extensive and time-consuming MD or MC simulations. It will be left for future investigations.
 
\subsection{Limiting cases without disorder}
In most cases one cannot solve the model presented above analytically. Using the quenched ensemble one cannot calculate the average over the disorder, while using the annealed ensemble one cannot obtain the partition sum after having performed the disorder average. Therefore we calculated sample averages for the quenched case above.  There are two special cases where the model can be solved exactly. These are the limiting cases where no disorder is present.  $\mathcal{P}=0$ is the situation of a normal Gaussian Chain with spring constant $\kappa$. It is well known that the mean square distance between two beads separated by $n$ monomers is given by $\left<R_n^2\right> = \frac{3}{\kappa} n$. The other limit, $\mathcal{P}=1$, corresponds to a fully connected network of beads. Assuming that all beads interact with spring constant $\kappa$, we can solve this problem analytically. Here we basically do not deal with a linear chain any more. The interaction matrix $K=(k_{ij})_{i,j}$ in this case writes
\begin{equation}
 k_{ij} = \begin{cases} N \kappa &\qquad \text{for}\quad i = j\\
			-\kappa &\qquad \text{for}\quad i\neq j
          \end{cases}
\end{equation}
By an easy calculation one can show that the inverse matrix is given by
\begin{equation}
 \sigma_{ij} = \begin{cases}
                \frac{2}{(N+1)\kappa} &\qquad \text{for}\quad i = j\\
		\frac{1}{(N+1)\kappa} &\qquad \text{for}\quad i \neq j
               \end{cases}
\end{equation}
Recall our definition of the chain at the beginning of sec.~\ref{sec:theory:genexp}: Although we have an $N\times N$-matrix our chain has $N+1$ beads, as we set $\mathbf{x}_0 \equiv 0$. Inserting into eq.~\eqref{eq:r2} yields
\begin{equation}
 \left< R_n^2\right> \equiv \left< r_{ij}^2\right> = \frac{3}{(N+1)\kappa/2}
\end{equation}
Within this system two beads are interacting with an effective harmonic potential with $\kappa_{\text{eff}} = (N+1)\kappa/2 $. 

\begin{figure}
 \includegraphics[angle=270, width=\hsize]{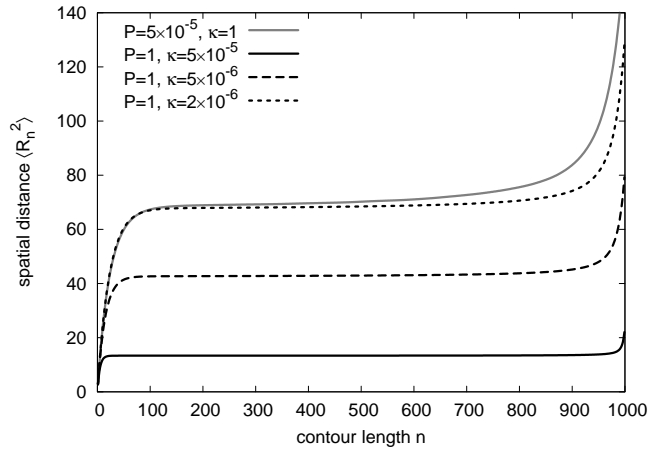}
\caption{\label{fig:meanfield} The Random Loop Model (RLM) compared to a system where the random attraction (setting $\kappa_{ij}=1$ with probability $\mathcal{P}$) has been replaced by an average attraction (setting all $\kappa_{ij}=\mathcal{P}\kappa$ for $|i-j|>1$). Shown are the RLM reference curve for $\mathcal{P}=5\times 10^{-5}$ (grey line), the corresponding system with average attraction (black line) and two systems with smaller average attraction.}
\end{figure}

Of major interest is the case where $\mathcal{P} = 1$, but where adjacent beads interact with a different spring constant than loops, i.e. $\kappa_{ij} = \kappa$ for $|i-j|=1$ and $\kappa_{ij} = \hat\kappa$ for $|i-j|>1$.  We were not able to solve this case analytically. One might take this system as a model for the low-temperature limit of the annealed case in eq.~\eqref{eq:V:ann:lt} where $\kappa$ is replaced by an effective interaction $\hat\kappa=\mathcal{P}\kappa$. On a more general footing this case might also be regarded as a model for a system where the random attraction with probability $\mathcal{P}$ and loop spring constant $\kappa$ has been replaced by an average attraction with probability $\mathcal{P}=1$ and loop spring constant $\mathcal{P}\kappa$. It is clear a priori that such a potential will lead to a collapse of the chain, as all beads are interconnected. In fig.~\ref{fig:meanfield} we chose $\hat\kappa=\kappa=1$ and $\mathcal{P}=4\times 10^{-5}$ as the reference curve. In comparison with the case of average attraction ($\mathcal{P}=1, \kappa=1, \hat\kappa=\mathcal{P}$) the levelling-off is much less pronounced. Of course it is possible to come into close agreement with the reference curve by choosing another interaction constant. For our reference curve one would have to lower $\hat\kappa$ by about one order of magnitude, corresponding to $\mathcal{P} \sim 2\times 10^{-6}$ ( $< 1$ loop per chain!). Although one could fit the data with these averaged attraction potential, we see no biological reason for such a potential to exist in the cell. 

\section{Conclusions}
In this paper we present a polymer model for the folding of the chromatin fibre in the interphase nucleus, based on recent experimental results~\cite{Mateos2007}. These show that the chromatin fibre inside the human cell nucleus is ruled by two different folding regimes: At small genomic distances, i.e. below $\sim$ 3 Mb the data can be explained well with a globular state polymer. At larger genomic distances there is a levelling-off to a scaling behaviour $\left<R^2\right>\sim O(1)$. This indicates the presence of long-range interactions. In agreement with the experimental findings we propose the existence of loops to be responsible for this levelling-off.  The main aspect of our model is the assumption of loops of various sizes and positions, where we average not only over the thermal ensemble but also over different configurations of loops. Keeping the looping probability $\mathcal{P}$ constant in the ensemble. Eq.~\eqref{eq:r2} gives a general expression for the mean square physical distance within the thermal ensemble between two arbitrary chosen beads, when each bead is allowed to interact with any other via harmonic potential.
We show that $\left<R^2\right>$ only depends on the interaction matrix $K$, which we have introduced in \eqref{eq:matrixK}. Within the Random Loop Model the matrix entries (equivalent to random loops) are chosen in a random manner to be $-\kappa$ with probability  $\mathcal{P}$ or $0$ otherwise. The average over the ensemble of loop configurations then turns out to be equivalent to averaging over the specific ensemble of random matrices with fixed $\mathcal{P}$. 
We show that random loops on all length scales explain the levelling-off observed in experiment, while restricting the loop sizes to only large loops (in the order of 50 - 100 Mb) or small loops (100 kb - 10 Mb) does not fit the data. 

In contrast to the Random-Walk/Giant-Loop model~\cite{Sachs1995} we do not assume fixed-size and regular placed loops. The average over the ensemble of different loop configurations turns out to be necessary to obtain the levelling-off observed in experiment. Its importance also becomes apparent in fig.~\ref{fig:cumulant}, where the RWGL model would yield the same result as the random walk, namely a constant: $c_4^{\text{RWGL}} = 5/3$.

In a recent paper~\cite{Mateos2007} we have shown that on short genomic distances the globular state fits the experimental data best compared to a self-avoiding and random walk. In the present study we have neglected the effect of excluded volume. Therefore on short genomic distances our model will explain the data inadequately as it shows random-walk-like behaviour by design on short contour lengths, because for a globular state polymer the exponent in eq.~\eqref{eq:scalinglaw} is $\nu=1/3$, while for a random walk polymer it is $\nu=1/2$.  On larger genomic distances the existence of a small number of loops can explain the levelling-off without the need for introducing excluded volume interactions. The role of excluded volume inside the cell nucleus on the spatial chromatin properties remains an open question, because of the impact of an enzyme called topoisomerase II, which is capable of cutting the DNA double strand in order to let another strand pass through it. This event might well give rise to a random-walk-like statistics, dependent on the frequency of these events. 

The model so far is able to explain some basic features of DNA folding revealed in recent experiments~\cite{Mateos2007}. Future work on this model might encompass for example the integration of excluded volume interactions and its effect on the mean square displacement. 
\begin{acknowledgements}
 M. Bohn gratefully acknowledges funding from the Landesgraduiertenf\"orderung Baden-W\"urttemberg.
\end{acknowledgements}


\begin{thebibliography}{23}
\expandafter\ifx\csname natexlab\endcsname\relax\def\natexlab#1{#1}\fi
\expandafter\ifx\csname bibnamefont\endcsname\relax
  \def\bibnamefont#1{#1}\fi
\expandafter\ifx\csname bibfnamefont\endcsname\relax
  \def\bibfnamefont#1{#1}\fi
\expandafter\ifx\csname citenamefont\endcsname\relax
  \def\citenamefont#1{#1}\fi
\expandafter\ifx\csname url\endcsname\relax
  \def\url#1{\texttt{#1}}\fi
\expandafter\ifx\csname urlprefix\endcsname\relax\def\urlprefix{URL }\fi
\providecommand{\bibinfo}[2]{#2}
\providecommand{\eprint}[2][]{\url{#2}}

\bibitem[{\citenamefont{Schiessel et~al.}(2001)\citenamefont{Schiessel,
  Gelbart, and Bruinsma}}]{Schiessel2001}
\bibinfo{author}{\bibfnamefont{H.}~\bibnamefont{Schiessel}},
  \bibinfo{author}{\bibfnamefont{W.~M.} \bibnamefont{Gelbart}},
  \bibnamefont{and} \bibinfo{author}{\bibfnamefont{R.}~\bibnamefont{Bruinsma}},
  \bibinfo{journal}{Biophys. J.} \textbf{\bibinfo{volume}{80}},
  \bibinfo{pages}{1940} (\bibinfo{year}{2001}).

\bibitem[{\citenamefont{Widom}(1989)}]{Widom1989}
\bibinfo{author}{\bibfnamefont{J.}~\bibnamefont{Widom}}, \bibinfo{journal}{Annu
  Rev Biophys Biophys Chem} \textbf{\bibinfo{volume}{18}}, \bibinfo{pages}{365}
  (\bibinfo{year}{1989}).

\bibitem[{\citenamefont{van Holde and Zlatanova}(1995)}]{Holde1995}
\bibinfo{author}{\bibfnamefont{K.}~\bibnamefont{van Holde}} \bibnamefont{and}
  \bibinfo{author}{\bibfnamefont{J.}~\bibnamefont{Zlatanova}},
  \bibinfo{journal}{J. Biol. Chem.} \textbf{\bibinfo{volume}{270}},
  \bibinfo{pages}{8373} (\bibinfo{year}{1995}).

\bibitem[{\citenamefont{van Holde and Zlatanova}(1996)}]{Holde1996}
\bibinfo{author}{\bibfnamefont{K.}~\bibnamefont{van Holde}} \bibnamefont{and}
  \bibinfo{author}{\bibfnamefont{J.}~\bibnamefont{Zlatanova}},
  \bibinfo{journal}{Proc. Natl. Acad. Sci. U. S. A.}
  \textbf{\bibinfo{volume}{93}}, \bibinfo{pages}{10548} (\bibinfo{year}{1996}).

\bibitem[{\citenamefont{Horowitz-Scherer and
  Woodcock}(2006)}]{Horowitz-Scherer2006}
\bibinfo{author}{\bibfnamefont{R.}~\bibnamefont{Horowitz-Scherer}}
  \bibnamefont{and} \bibinfo{author}{\bibfnamefont{C.}~\bibnamefont{Woodcock}},
  \bibinfo{journal}{Chromosoma} \textbf{\bibinfo{volume}{115}},
  \bibinfo{pages}{1} (\bibinfo{year}{2006}).

\bibitem[{\citenamefont{Mateos-Langerak
  et~al.}(2007)\citenamefont{Mateos-Langerak, Giromus, de~Leeuw, Bohn, Kreth,
  Heermann, van Driel, and Goetze}}]{Mateos2007}
\bibinfo{author}{\bibfnamefont{J.}~\bibnamefont{Mateos-Langerak}},
  \bibinfo{author}{\bibfnamefont{O.}~\bibnamefont{Giromus}},
  \bibinfo{author}{\bibfnamefont{W.}~\bibnamefont{de~Leeuw}},
  \bibinfo{author}{\bibfnamefont{M.}~\bibnamefont{Bohn}},
  \bibinfo{author}{\bibfnamefont{G.}~\bibnamefont{Kreth}},
  \bibinfo{author}{\bibfnamefont{D.~W.} \bibnamefont{Heermann}},
  \bibinfo{author}{\bibfnamefont{R.}~\bibnamefont{van Driel}},
  \bibnamefont{and} \bibinfo{author}{\bibfnamefont{S.}~\bibnamefont{Goetze}},
  \bibinfo{journal}{arXiv:0705.1656 [q-bio.GN]}  (\bibinfo{year}{2007}).

\bibitem[{\citenamefont{Hahnfeldt et~al.}(1993)\citenamefont{Hahnfeldt, Hearst,
  Brenner, Sachs, and Hlatky}}]{Hahnfeldt1993}
\bibinfo{author}{\bibfnamefont{P.}~\bibnamefont{Hahnfeldt}},
  \bibinfo{author}{\bibfnamefont{J.~E.} \bibnamefont{Hearst}},
  \bibinfo{author}{\bibfnamefont{D.~J.} \bibnamefont{Brenner}},
  \bibinfo{author}{\bibfnamefont{R.~K.} \bibnamefont{Sachs}}, \bibnamefont{and}
  \bibinfo{author}{\bibfnamefont{L.~R.} \bibnamefont{Hlatky}},
  \bibinfo{journal}{Proc. Natl. Acad. Sci. U. S. A.}
  \textbf{\bibinfo{volume}{90}}, \bibinfo{pages}{7854} (\bibinfo{year}{1993}).

\bibitem[{\citenamefont{Yokota et~al.}(1995)\citenamefont{Yokota, van~den Engh,
  Hearst, Sachs, and Trask}}]{Yokota1995}
\bibinfo{author}{\bibfnamefont{H.}~\bibnamefont{Yokota}},
  \bibinfo{author}{\bibfnamefont{G.}~\bibnamefont{van~den Engh}},
  \bibinfo{author}{\bibfnamefont{J.}~\bibnamefont{Hearst}},
  \bibinfo{author}{\bibfnamefont{R.}~\bibnamefont{Sachs}}, \bibnamefont{and}
  \bibinfo{author}{\bibfnamefont{B.}~\bibnamefont{Trask}}, \bibinfo{journal}{J.
  Cell Biol.} \textbf{\bibinfo{volume}{130}}, \bibinfo{pages}{1239}
  (\bibinfo{year}{1995}).

\bibitem[{\citenamefont{Sachs et~al.}(1995)\citenamefont{Sachs, Engh, Trask,
  Yokota, and Hearst}}]{Sachs1995}
\bibinfo{author}{\bibfnamefont{R.}~\bibnamefont{Sachs}},
  \bibinfo{author}{\bibfnamefont{G.}~\bibnamefont{Engh}},
  \bibinfo{author}{\bibfnamefont{B.}~\bibnamefont{Trask}},
  \bibinfo{author}{\bibfnamefont{H.}~\bibnamefont{Yokota}}, \bibnamefont{and}
  \bibinfo{author}{\bibfnamefont{J.}~\bibnamefont{Hearst}},
  \bibinfo{journal}{Proc. Natl. Acad. Sci. U. S. A.}
  \textbf{\bibinfo{volume}{92}}, \bibinfo{pages}{2710} (\bibinfo{year}{1995}).

\bibitem[{\citenamefont{M\"unkel and Langowski}(1998)}]{Munkel1998}
\bibinfo{author}{\bibfnamefont{C.}~\bibnamefont{M\"unkel}} \bibnamefont{and}
  \bibinfo{author}{\bibfnamefont{J.}~\bibnamefont{Langowski}},
  \bibinfo{journal}{Phys. Rev. E} \textbf{\bibinfo{volume}{57}},
  \bibinfo{pages}{5888} (\bibinfo{year}{1998}).

\bibitem[{\citenamefont{M\"unkel et~al.}(1999)\citenamefont{M\"unkel, Eils,
  Dietzel, Zink, Mehring, Wedemann, Cremer, and Langowski}}]{Munkel1999}
\bibinfo{author}{\bibfnamefont{C.}~\bibnamefont{M\"unkel}},
  \bibinfo{author}{\bibfnamefont{R.}~\bibnamefont{Eils}},
  \bibinfo{author}{\bibfnamefont{S.}~\bibnamefont{Dietzel}},
  \bibinfo{author}{\bibfnamefont{D.}~\bibnamefont{Zink}},
  \bibinfo{author}{\bibfnamefont{C.}~\bibnamefont{Mehring}},
  \bibinfo{author}{\bibfnamefont{G.}~\bibnamefont{Wedemann}},
  \bibinfo{author}{\bibfnamefont{T.}~\bibnamefont{Cremer}}, \bibnamefont{and}
  \bibinfo{author}{\bibfnamefont{J.}~\bibnamefont{Langowski}},
  \bibinfo{journal}{J. Mol. Biol.} \textbf{\bibinfo{volume}{285}},
  \bibinfo{pages}{1053} (\bibinfo{year}{1999}).

\bibitem[{\citenamefont{Palstra et~al.}(2003)\citenamefont{Palstra, Tolhuis,
  Splinter, Nijmeijer, Grosveld, and de~Laat}}]{Palstra2003}
\bibinfo{author}{\bibfnamefont{R.-J.} \bibnamefont{Palstra}},
  \bibinfo{author}{\bibfnamefont{B.}~\bibnamefont{Tolhuis}},
  \bibinfo{author}{\bibfnamefont{E.}~\bibnamefont{Splinter}},
  \bibinfo{author}{\bibfnamefont{R.}~\bibnamefont{Nijmeijer}},
  \bibinfo{author}{\bibfnamefont{F.}~\bibnamefont{Grosveld}}, \bibnamefont{and}
  \bibinfo{author}{\bibfnamefont{W.}~\bibnamefont{de~Laat}},
  \bibinfo{journal}{Nat. Genet.} \textbf{\bibinfo{volume}{35}},
  \bibinfo{pages}{190} (\bibinfo{year}{2003}).

\bibitem[{\citenamefont{de~Laat and Grosveld}(2003)}]{Laat2003}
\bibinfo{author}{\bibfnamefont{W.}~\bibnamefont{de~Laat}} \bibnamefont{and}
  \bibinfo{author}{\bibfnamefont{F.}~\bibnamefont{Grosveld}},
  \bibinfo{journal}{Chromosome Res.} \textbf{\bibinfo{volume}{11}},
  \bibinfo{pages}{447} (\bibinfo{year}{2003}).

\bibitem[{\citenamefont{Petrascheck et~al.}(2005)\citenamefont{Petrascheck,
  Escher, Mahmoudi, Verrijzer, Schaffner, and Barberis}}]{Petrascheck2005}
\bibinfo{author}{\bibfnamefont{M.}~\bibnamefont{Petrascheck}},
  \bibinfo{author}{\bibfnamefont{D.}~\bibnamefont{Escher}},
  \bibinfo{author}{\bibfnamefont{T.}~\bibnamefont{Mahmoudi}},
  \bibinfo{author}{\bibfnamefont{C.~P.} \bibnamefont{Verrijzer}},
  \bibinfo{author}{\bibfnamefont{W.}~\bibnamefont{Schaffner}},
  \bibnamefont{and} \bibinfo{author}{\bibfnamefont{A.}~\bibnamefont{Barberis}},
  \bibinfo{journal}{Nucleic Acids Res.} \textbf{\bibinfo{volume}{33}},
  \bibinfo{pages}{3743} (\bibinfo{year}{2005}).

\bibitem[{\citenamefont{Fraser}(2006)}]{Fraser2006}
\bibinfo{author}{\bibfnamefont{P.}~\bibnamefont{Fraser}},
  \bibinfo{journal}{Curr. Opin. Genet. Dev.} \textbf{\bibinfo{volume}{16}},
  \bibinfo{pages}{490} (\bibinfo{year}{2006}).

\bibitem[{\citenamefont{Fraser and Bickmore}(2007)}]{Fraser2007}
\bibinfo{author}{\bibfnamefont{P.}~\bibnamefont{Fraser}} \bibnamefont{and}
  \bibinfo{author}{\bibfnamefont{W.}~\bibnamefont{Bickmore}},
  \bibinfo{journal}{Nature} \textbf{\bibinfo{volume}{447}},
  \bibinfo{pages}{413} (\bibinfo{year}{2007}).

\bibitem[{\citenamefont{Cook}(2002)}]{Cook2002}
\bibinfo{author}{\bibfnamefont{P.~R.} \bibnamefont{Cook}},
  \bibinfo{journal}{Nat. Genet.} \textbf{\bibinfo{volume}{32}},
  \bibinfo{pages}{347} (\bibinfo{year}{2002}).

\bibitem[{\citenamefont{Grosberg and Khokhlov}(1994)}]{Grosberg1994}
\bibinfo{author}{\bibfnamefont{A.~Y.} \bibnamefont{Grosberg}} \bibnamefont{and}
  \bibinfo{author}{\bibfnamefont{A.~R.} \bibnamefont{Khokhlov}},
  \emph{\bibinfo{title}{Statistical Physics of Macromolecules}}
  (\bibinfo{publisher}{AIP Press}, \bibinfo{year}{1994}).

\bibitem[{\citenamefont{de~Gennes}(1979)}]{Gennes1979}
\bibinfo{author}{\bibfnamefont{P.-G.} \bibnamefont{de~Gennes}},
  \emph{\bibinfo{title}{Scaling concepts in polymer physics}}
  (\bibinfo{publisher}{Ithaca, N.Y., Cornell University Press},
  \bibinfo{year}{1979}).

\bibitem[{\citenamefont{Heermann and Bohn}(2007)}]{Heermann2007}
\bibinfo{author}{\bibfnamefont{D.~W.} \bibnamefont{Heermann}} \bibnamefont{and}
  \bibinfo{author}{\bibfnamefont{M.}~\bibnamefont{Bohn}},
  \bibinfo{journal}{arXiv:0705.1241v1 [cond-mat.stat-mech]}
  (\bibinfo{year}{2007}).

\bibitem[{\citenamefont{Baiesi et~al.}(2003)\citenamefont{Baiesi, Carlon,
  Kafri, Mukamel, Orlandini, and Stella}}]{Baiesi2003}
\bibinfo{author}{\bibfnamefont{M.}~\bibnamefont{Baiesi}},
  \bibinfo{author}{\bibfnamefont{E.}~\bibnamefont{Carlon}},
  \bibinfo{author}{\bibfnamefont{Y.}~\bibnamefont{Kafri}},
  \bibinfo{author}{\bibfnamefont{D.}~\bibnamefont{Mukamel}},
  \bibinfo{author}{\bibfnamefont{E.}~\bibnamefont{Orlandini}},
  \bibnamefont{and} \bibinfo{author}{\bibfnamefont{A.~L.}
  \bibnamefont{Stella}}, \bibinfo{journal}{Phys. Rev. E}
  \textbf{\bibinfo{volume}{67}}, \bibinfo{pages}{021911}
  (\bibinfo{year}{2003}).

\bibitem[{\citenamefont{Fisher}(1966)}]{Fisher1966}
\bibinfo{author}{\bibfnamefont{M.}~\bibnamefont{Fisher}}, \bibinfo{journal}{J.
  Chem. Phys.} \textbf{\bibinfo{volume}{44}}, \bibinfo{pages}{616}
  (\bibinfo{year}{1966}).

\bibitem[{\citenamefont{Binder and Young}(1986)}]{Binder1986}
\bibinfo{author}{\bibfnamefont{K.}~\bibnamefont{Binder}} \bibnamefont{and}
  \bibinfo{author}{\bibfnamefont{A.~P.} \bibnamefont{Young}},
  \bibinfo{journal}{Rev. Mod. Phys.} \textbf{\bibinfo{volume}{58}},
  \bibinfo{pages}{801} (\bibinfo{year}{1986}).

\end{thebibliography}

\end{document}